%% file: bandex.tex
\documentclass[10pt,times,twocolumn]{article}
\usepackage{epsfig}
\usepackage{bar}
\usepackage{times}
\usepackage{latex8}

\newcommand{\comment}[1]{}

\comment{
\textwidth 6.5in
\oddsidemargin 0in
\evensidemargin 0in
\columnsep 0.25in
\topmargin -0.5in
\textheight 9.0in
}

\comment{

\makeatletter
\def\@normalsize{\@setsize\normalsize{12pt}\xpt\@xpt
\abovedisplayskip 10pt plus2pt minus5pt\belowdisplayskip \abovedisplayskip
\abovedisplayshortskip \z@ plus3pt\belowdisplayshortskip 6pt plus3pt
minus3pt\let\@listi\@listI}
   
\def\subsize{\@setsize\subsize{12pt}\xipt\@xipt}
    
\def\section{\@startsection {section}{1}{\z@}{24pt plus 2pt minus 2pt}
{12pt plus 2pt minus 2pt}{\large\bf}}
 
\def\subsection{\@startsection {subsection}{2}{\z@}{12pt plus 2pt minus 2pt}
{12pt plus 2pt minus 2pt}{\subsize\bf}}
\makeatother                                                                 

}

\newcommand{\smallcaption}[1]{\caption[#1]{{\protect\small \protect\bf #1}}}

\newcommand{\bx}{{\sc Band-X}}

\begin{document}

\title{The Bandwidth Exchange Architecture}

\renewcommand{\baselinestretch}{1}
\author{
	David~Michael~Turner\\
	Dept. of Computer Science\\
	Drexel University\\
	{\em dmt36@drexel.edu}
\and
	Vassilis~Prevelakis\\
	Dept. of Computer Science\\
	Drexel University\\
	{\em vp@cs.drexel.edu}
\and
	Angelos~D.~Keromytis\\
	Dept. of Computer Science\\
	Columbia University\\
	{\em angelos@cs.columbia.edu}
}


\maketitle
\thispagestyle{empty}

\input{01-abstract}
\input{10-introduction}

\input{20-architecture}

\input{40-implementation}

\input{90-related}
\input{95-conclusions}

\bibliographystyle{latex8}
\bibliography{bandex}
\end{document}

%% file: 01-abstract.tex
%
%

\begin{abstract}
\label{sec:abstract}

New applications for the Internet such as video on demand, grid
computing {\it etc.} depend on the availability of high bandwidth
connections with acceptable Quality of Service (QoS). There appears to
be, therefore, a requirement for a market where bandwidth-related
transactions can take place.  For this market to be effective, it must
be efficient for both the provider (seller) and the user (buyer) of
the bandwidth. This implies that: \emph{(a)} the buyer must have a
wide choice of providers that operate in a competitive environment,
\emph{(b)} the seller must be assured that a QoS transaction will be
paid by the customer, and \emph{(c)} the QoS transaction establishment
must have low overheads so that it may be used by individual customers
without a significant burden to the provider.

In order to satisfy these requirements, we propose a framework that
allows customers to purchase bandwidth using an open market where
providers advertise links and capacities and customers bid for these
services. The model is close to that of a commodities market that
offers both advance bookings (futures) and a spot market. We explore
the mechanisms that can support such a model.

\comment{
\vskip 0.1in
\noindent {\bf Conference category:} Research
}
\end{abstract}

%% file: 10-introduction.tex
%
%

\Section{Introduction}
\label{sec:intro}

Years of research on Quality of Service (QoS) architectures for the
Internet have resulted in sophisticated proposals that have not been
broadly exploited commercially. In particular, Integrated Services
(IntServ) \cite{rfc-rsvp} and Differentiated Services (DiffServ)
\cite{diffserv:rfc} have long been supported by major router and
operating system vendors, yet have only seen minimal use in
practice. One explanation offered by the networking and QoS community
has been a lack of a commercialization model,
together with the necessary accounting and charging architecture
\cite{payer2003}. A related crucial issue is assurance of end-to-end
QoS coherence in the face of multiple intervening parties, such as
transit ISPs.

These two issues, taken together, are responsible for suppressing
interest from both the ISPs (in commercially exploiting QoS to its
full potential) and the users (in taking advantage of such
services). Simply put, if an ISP cannot be paid for reserving
bandwidth to a user, they will not offer QoS; if users cannot be
assured of end-to-end QoS, they will not pay for the
service. Compounding the problem is the issue of management: it is
certainly possible for a large entity, such as a multi-national
company, to coordinate with the relevant ISPs so that its various
geographically dispersed networks are connected provisioned using a
series of DiffServ or IntServ tunnels. However, the effort is
considerable and requires manual intervention from a number of
people. Perhaps most importantly, the ISPs' network operations centers
(NOCs) will need to configure the various routers
appropriately. Clearly, such an approach will not scale well if
preferentially treated bandwidth is to become a commodity that can be
traded, as has been recognized before \cite{davie2003}.  Yet, the
increasing use of the Internet for time-sensitive or otherwise
critical applications effectively mandate some form of bandwidth
reservation, often for short periods of time ({\it e.g.,} watching
a movie).

We present a market-based approach to self-managing QoS across
multiple ISPs. Our architecture introduces a Bandwidth Exchange (\bx),
which facilitates the trading of reserved bandwidth between ISPs and
users. This facility allows purchasing bandwidth in advance (effectively
creating a ``futures'' market for bandwidth) as well as on the
``spot'' market. Users can select from a range of offerings by various
ISPs to create an end-to-end pipe (with the desired bandwidth and QoS)
piece-meal, or can choose to
purchase a complete package from a single provider (or consortium of
providers), where available. This is similar to the way people
purchase low-cost airplane tickets online.

To ease the task of accounting and administration, we use the
micropayment architecture introduced in \cite{sodapaper} to provide
both accounting and authorization. Briefly, users purchasing bandwidth
on \bx\ are provided with credentials that allow them to establish the
necessary QoS pipes among the necessary network elements (routers),
within the constraints of their contracts. Our use of a
trust-management system (KeyNote \cite{KeyNote}) allows us to perform
both billing and authorization with the same mechanism, simplifying
the architecture and eliminating the need for manual configuration or
universal trust of the \bx\ service ({\it e.g.,} to configure the
relevant routers of several ISPs).

To better illustrate the use of the \bx\ architecture, we next
describe a sample usage scenario involving an end user and several
ISPs. In Section~\ref{sec:arch} we present the system architecture in
more detail. Section~\ref{sec:implementation} describes the various
components of our system, in particular our micro-checks mechanism, and
how they operate together, along with a security analysis. We discuss
related work in Section~\ref{sec:rel}.

\SubSection{Motivation}

Consider the following scenario of a user Alice wishing to reserve an
end-to-end 50Mbps ``pipe'' from Rome to Dublin\footnote{We use
geographical identifiers instead of IP addresses to simplify the
example.}. Using an appropriate tool ({\it e.g.,} auction site,
database, service bureau) she decides to purchase a link from Rome to
Paris offered by ISP A, and another link from Paris to Dublin offered by
ISP B. However, Alice does not need the QoS pipe immediately; rather,
she needs it for the time her remote presentation is scheduled, a
few days later.

Payment may be effected in various ways (examples given later in the
paper) depending on the policy of each ISP. Once the reservation has
been booked, each ISP sends a credential to Alice authorizing her to
use the required link at the desired time and date and for the
appropriate time interval. The credentials are set to expire at the
end of the reserved period. Again, depending on the way payment is
handled and the policies of the ISPs and other involved parties, more
than these two credentials may be required for access to be granted
(this is explained later).

Just before the link needs to be established, Alice's QoS negotiation
agent (QNA) will send a QoS request to the network elements (NEs) of
the two ISPs to ensure that the appropriate resources have been
allocated. Since two providers are involved, Alice's QNA will need to
contact each ISP separately. Depending on the bandwidth reservation
protocol used, Alice's QNA may communicate with a central entity
within the ISP, or may negotiate a path through the ISP's network and
then reserve the desired bandwidth with each network element separately.

For this discussion, we have limited ourselves to bandwidth
reservation; additional QoS requirements (such as latency) may be
specified within the same framework.


\noindent {\bf Spot Market} \hskip .1in
Given an efficient purchasing mechanism, an ``advance'' booking such
as the one mentioned earlier may be made even seconds before the
channel will be used, so the term ``spot market'' is used to define a
different payment regime that may be used to sell the unused network
capacity. The ``spot market'' allows premium best-effort services to
be sold. In this case, we are not making any promises regarding
availability of bandwidth, but we say that by paying a small premium,
packets may be treated favorably in the allocation of the remaining
bandwidth (after the booked commitments are served).

%% file: 20-architecture.tex
%
%

\newlength{\credwidth}

\Section{Architecture}
\label{sec:arch}


\SubSection{Operation of the Spot Market}
\label{ss:spotOv}

\begin{figure}
\begin{center}
\epsfig{file=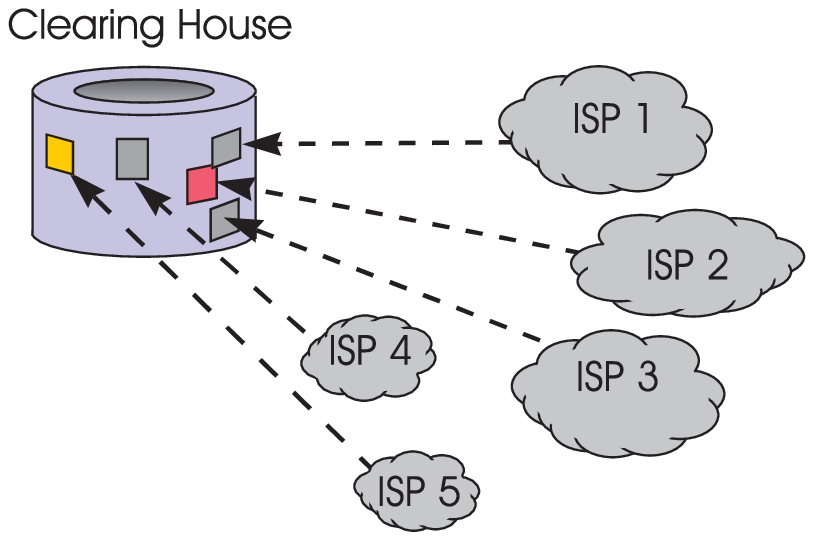,width=2.2in}
\end{center}
\smallcaption{The \bx\ Clearing House acts as a repository of all the
offers for bandwidth issued by the ISPs.}
\label{fig:T01PostOffer}
\end{figure}

Initially, the various bandwidth providers post their available
capacities in the \bx\ clearing house. The system can accommodate one
or more such clearing houses, since they function as announcement
boards. Apart from that, the clearing house is not involved in the
purchase of bandwidth (see Figure~\ref{fig:T01PostOffer}).

The postings are of the form of credentials that describe the identity
of the ISP and promise to abide by a set of QoS specifications between
two points of the ISPs network. The credential may also contain the
time period that the offer is valid (which may be different from the
expiration of the credential), the price of the concession, and
additional ISP-related information, such as the path that should be
taken between the two points. Offer credentials are signed by the ISP
who issues them.

Customers contact the Clearing House to collect offers from the
ISPs. For complex paths, a customer may need to collect more than one
offer and use them together. In an environment with a single clearing
house, the customer can issue queries to get lists of offers matching
his or her requirements.  If there are many clearing houses, the
customer may dispatch an intelligent agent to collect the offers and
come back with a recommendation that meets preassigned constraints
(price, ISP reliability {\it etc.}), query each clearing house
independently, or use a meta-search engine.

\begin{figure}
\begin{minipage}[b]{0.45\textwidth}
\begin{center}
\epsfig{file=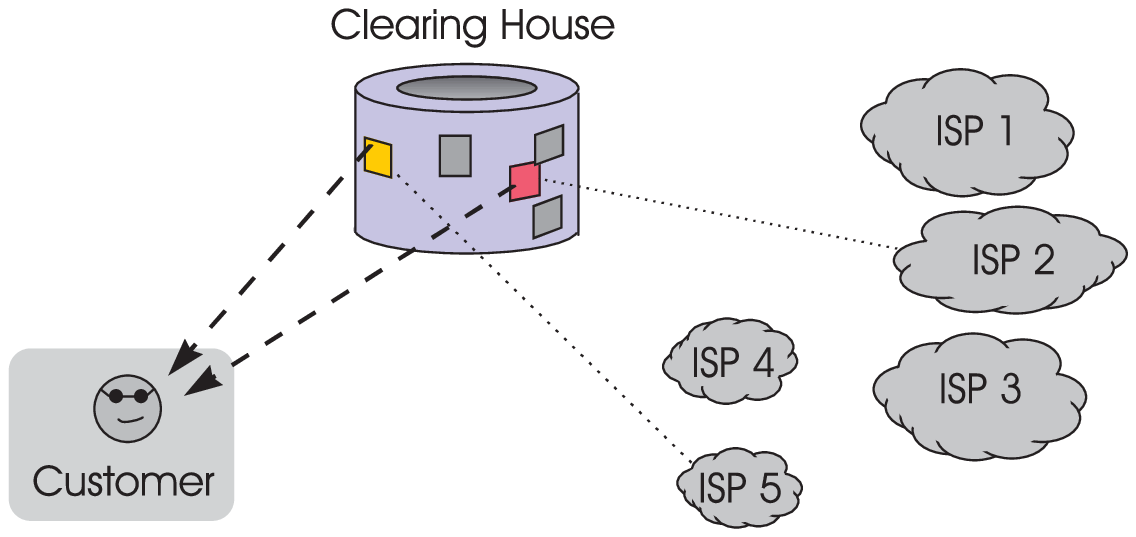,width=2.9in}
\end{center}
\smallcaption{Customer finalizes the path selection by downloading the
offer credentials.}
\label{fig:T02CustSel}
\end{minipage}
\hskip .3in
\begin{minipage}[b]{0.5\textwidth}
\begin{center}
\epsfig{file=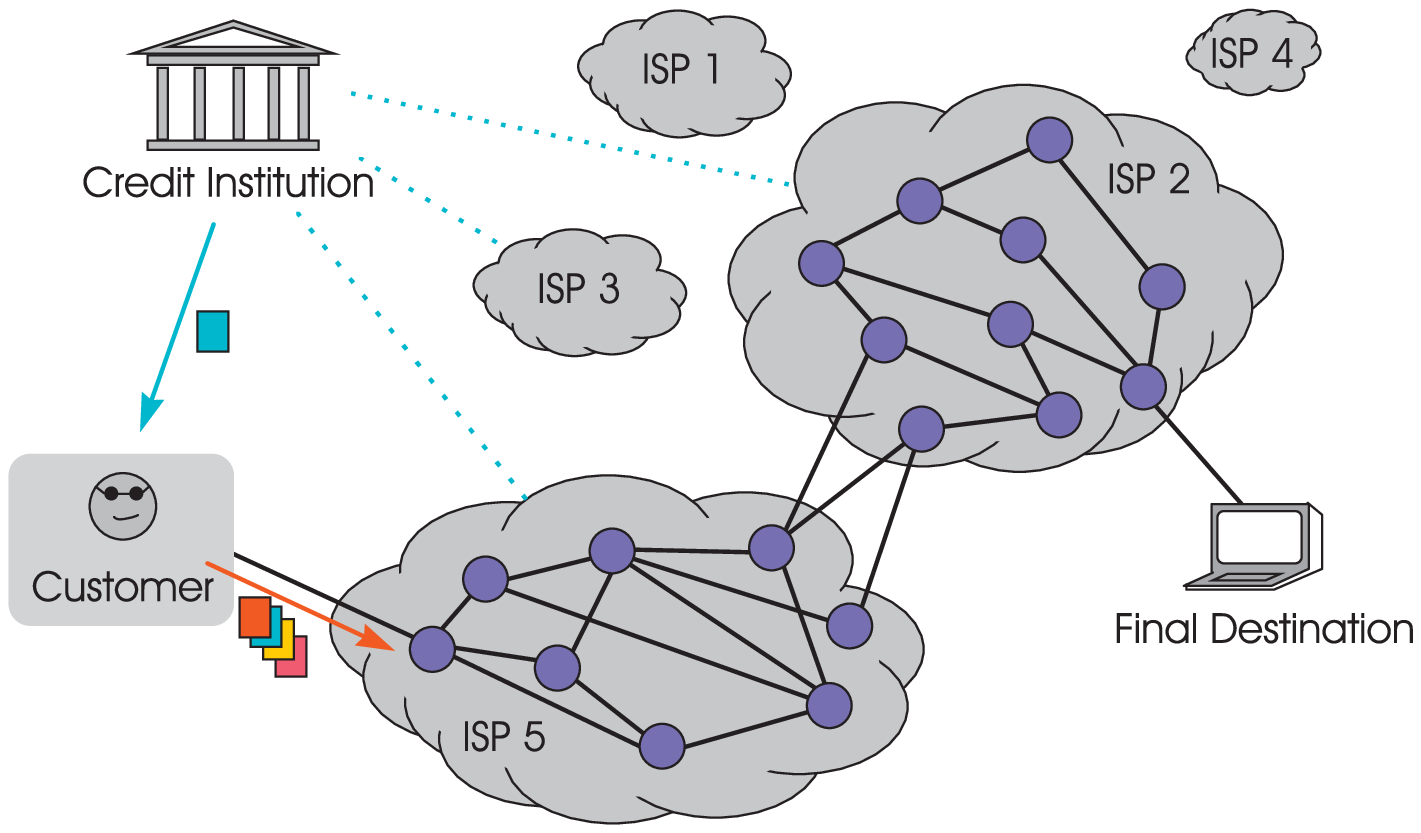,width=2.9in}
\end{center}
\smallcaption{The customer issues a reservation request by sending the
offer credentials collected from the \bx\ Clearing House along with a
credit-worthiness credential issued by his or her credit institution.}
\label{fig:T03CustPay}
\end{minipage}
\end{figure}

At the end of the search, the customer will hold one or more offer
credentials that describe the desired path and QoS specs, as shown in
Figure~\ref{fig:T02CustSel}.

At this point, the customer has not actually purchased the
bandwidth. In order to issue payment and reserve the bandwidth, a
number of steps have to be taken.  The customer (or the host at one of
the end-points of the connection) contacts the first-hop network
element (NE) and activates the reservation protocol. The NE issues a
challenge which is then returned signed by the customer. This response
also contains the offer credentials collected by the customer and a
credit-worthiness credential issued by the customer's credit
institution, as shown in Figure~\ref{fig:T03CustPay}.

\begin{figure}
\begin{minipage}[b]{0.47\textwidth}
\begin{center}
\epsfig{file=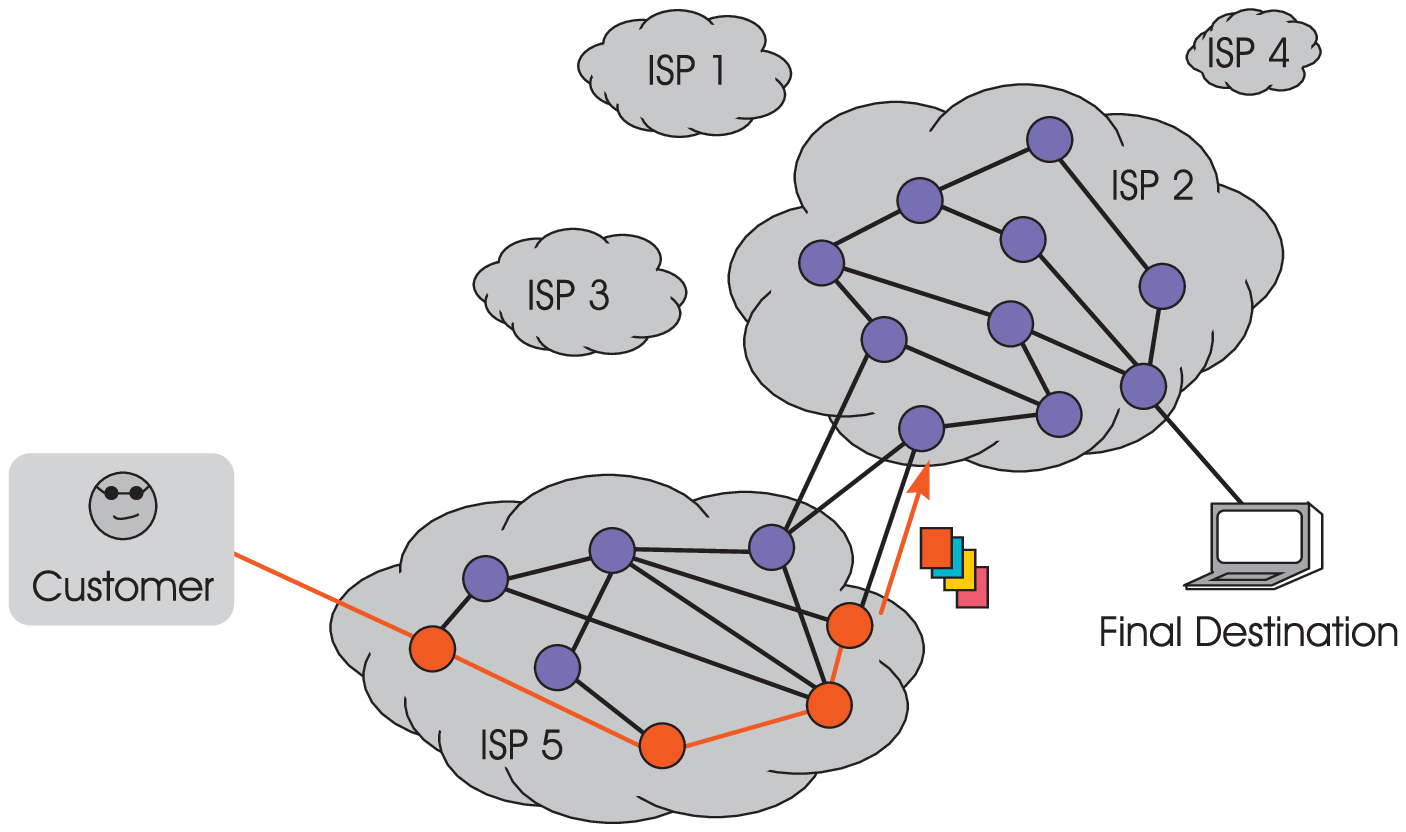,width=2.9in}
\end{center}
\smallcaption{Each time the path crosses ISP boundaries, additional
negotiations have to be carried out, to ensure that the next-hop ISP
can be paid for passage.}
\label{fig:T04InterISP}
\end{minipage}
\hskip .3in
\begin{minipage}[b]{0.48\textwidth}
\begin{center}
\epsfig{file=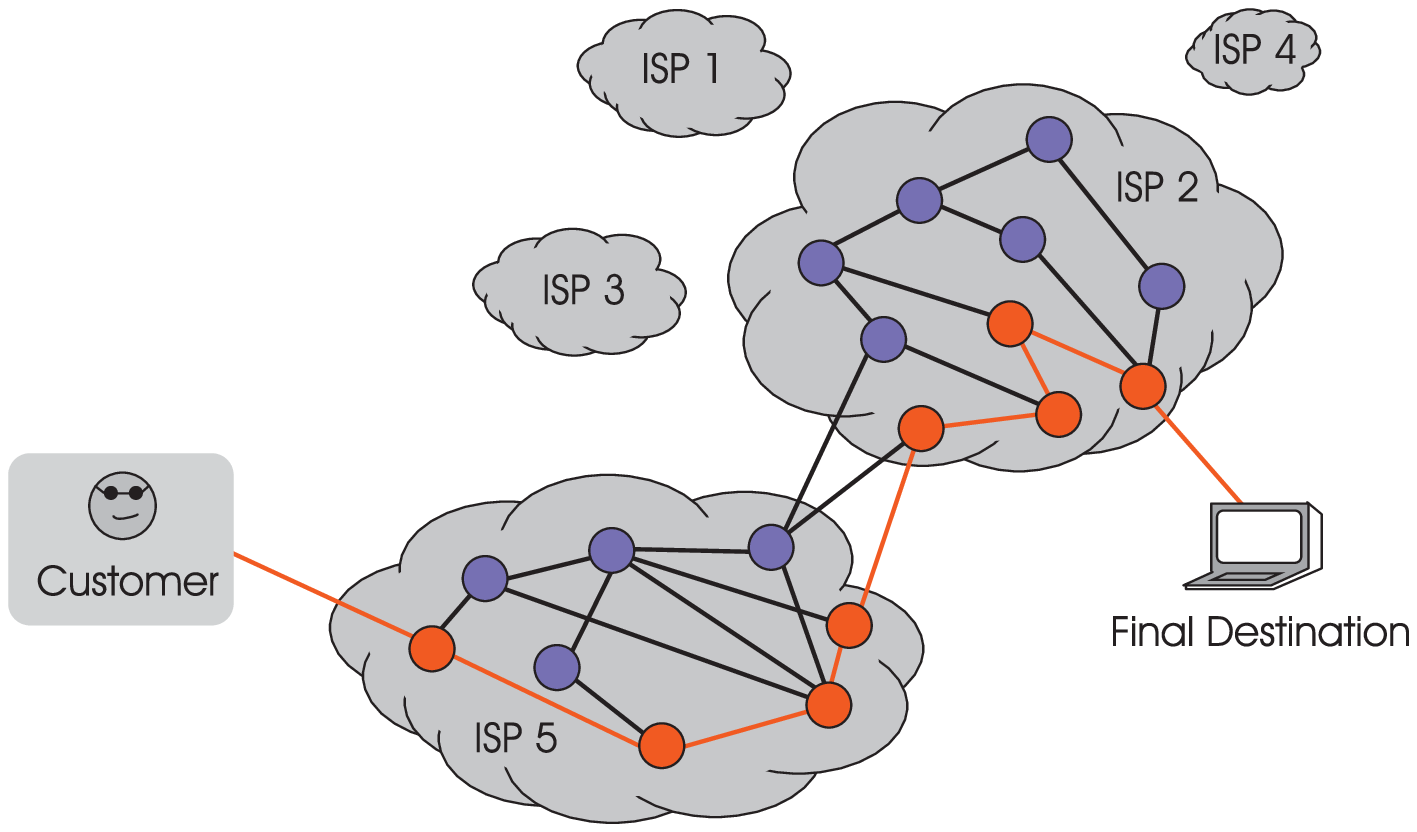,width=2.9in}
\end{center}
\smallcaption{The path has now been established and communication can
proceed.}
\label{fig:T05Success}
\end{minipage}
\end{figure}

This exchange accomplishes the following: $(a)$ identifies the
customer (the key that has signed the NE challenge), $(b)$ provides
proof of good standing (the credential issued by the credit
institution to the customer's key), $(c)$ limits payment only to the
offer credentials provided, $(d)$ can be used only for that particular
transaction since it depends on the challenge issued by the NE.  On
the basis of this transaction, the first hop NE contacts other NEs
within the ISPs network establishing the purchased path. If the path
crosses ISP boundaries, additional transactions have to be carried out
between the NE of the new ISP and the end user, as shown in
Figure~\ref{fig:T04InterISP}.

When the last hop is reached, the connection is considered established
and the final destination host can initiate a connection with the
customer's host over the reserved path (Figure~\ref{fig:T05Success}).

There is no need for the ISPs offers to match exactly the requirements
of the customer. For example, if Alice requires a 50Mbps link from
Atlanta to Dublin, she may use an offer for a 100Mbps connection, but
purchase only 50Mbps. The providers may include clauses in their offer
credentials allowing or prohibiting such un-bundling. The flexibility
of the policy language used in \bx\ allows many such special
considerations to be encoded within the offer credentials. The
advantage of having these restrictions expressed as policy is that
they can be used directly by the ISP's infrastructure without any need
for conversion. Moreover, the customer cannot alter these restrictions
since they are an integral part of the credential (and are protected
by the ISP's signing of the offer credentials).

\SubSection{Operation of the Futures Market}
\label{ss:futuresOv}

In the Spot Market, the customer collects the offers and sets up the
path in short order, because the offers are effective immediately and
have a short lifetime. There is no need to negotiate with the ISPs
before the reservation.

In the Futures Market the situation is different, since the ISPs need
to know what bandwidth has been purchased to plan their resource
allocation.  Once the customer collects the offers, a notional
reservation negotiation will be initiated. The negotiation is notional
because no state changes are actually effected on the network
elements. The customer's QNA will not detect any change in the
negotiation. Within the ISPs network, no path is created; rather the
reservation is entered in the ISP's database, and a reservation
credential is sent to the end user. This credential will then be used
in the same manner as the offer credential was used in the Spot Market
scenario. Since the bandwidth has been paid for, the reservation
credential commits the ISPs to provide the requested resources at the
appropriate future time.

At that time (when the path is actually required) the customer
initiates a reservation negotiation, but sends only the reservation
credential (instead of the offer and credit institution
credentials). The ISP network elements will reserve the path as
specified in the reservation credential. The case of multiple ISPs is
handled in a similar manner.

\SubSection{Role of the Credit Institution}
\label{ss:CreditInst}

Like the Clearing House, there is no requirement to have a single
Credit Institution. It is, however, important that the ISPs have a way
of confirming the keys of the various Credit Institutions. This is
because the credit-worthiness credentials (CWCs) issued by the Credit
Institutions to their customers will have to be verified by each
ISP. If an ISP cannot verify a CWC, then it may be fake; trusting it
may result in the equivalent of a bounced check.

%% file: 40-implementation.tex
\Section{Implementation}
\label{sec:implementation}

\comment{
	Here, we describe some of the components in our architecture --- in
	particular, the micropayment mechanism. We provide an explanation of
	its operation, and a short security analysis.
}

\SubSection{KeyNote Microchecks}
\label{ss:microchecks}

\begin{figure}[ht]
\center{\epsfig{file=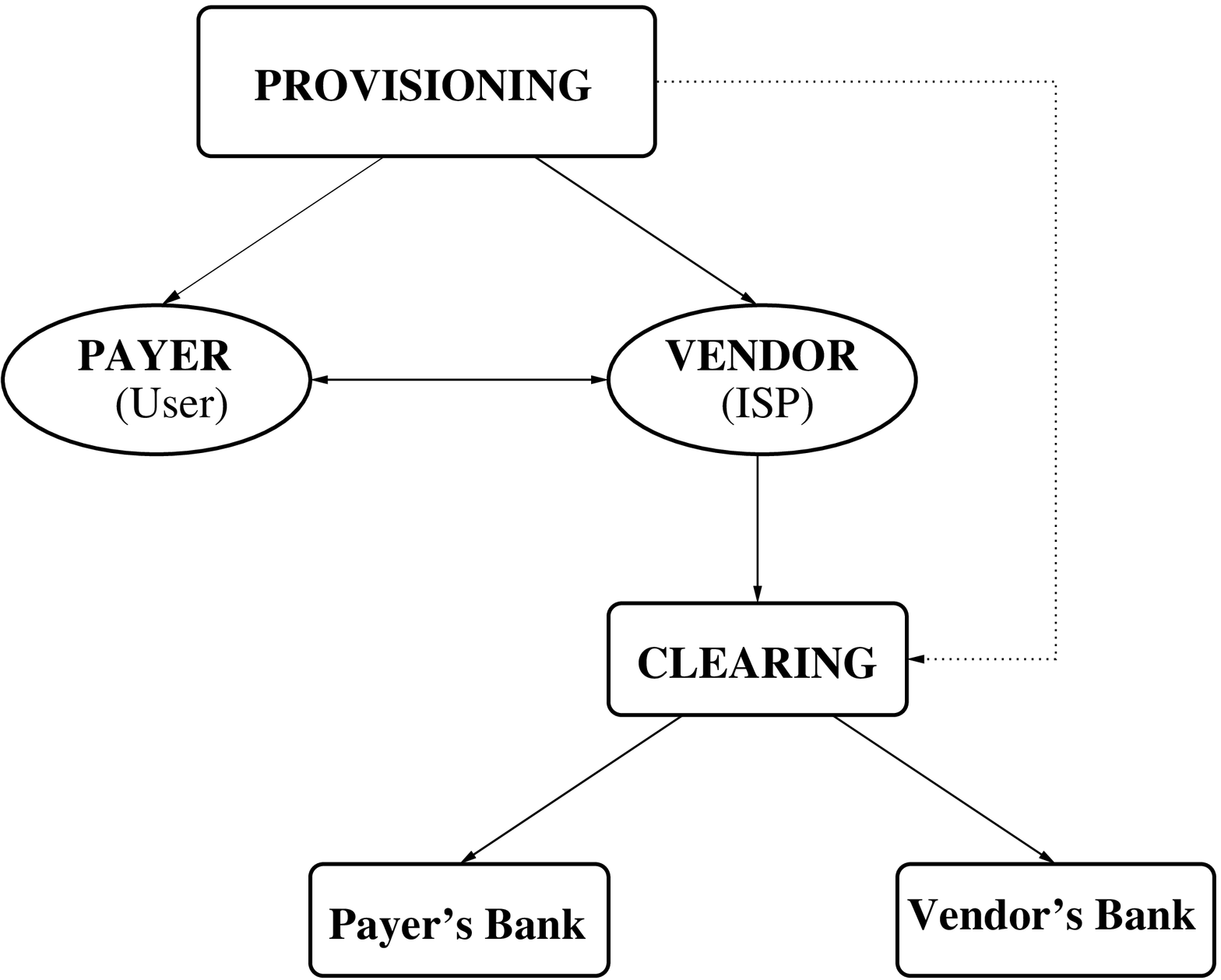,width=2.5in}}
\smallcaption{Microbilling architecture diagram. We have the generic terms
for each component, and in parentheses the corresponding players in
\bx. The arrows represent communication between the two parties:
Provisioning issues credentials to Payers and Merchants; these
communicate to complete transactions; Merchants send transaction
information to Clearing which verifies the transaction and posts the
necessary credits/charges or arranges money transfers. Provisioning
and Clearing exchange information on the status of Payer and Merchant
accounts.}
\label{figure:workflow}
\end{figure}

The micro-payments system introduced in \cite{sodapaper} forms the basis
of our approach. The general architecture of this micro-billing system
is shown in Figure~\ref{figure:workflow}.  Under \bx, a Merchant is an
ISP selling bandwidth and a Payer is a client wishing to make a QoS
reservation.

In this system, Provisioning issues KeyNote \cite{KeyNote}
credentials to users (Payers) and ISPs (Merchants). These credentials
describe the conditions under which a user is allowed to perform a
transaction ({\it i.e.,} the user's credit limit) and the fact that a
Merchant is authorized to participate in a particular transaction.

Initially, the ISP encodes the details of the available bandwidth into
an {\em offer} which is uploaded to the \bx\ site, along with a
credential that authorizing any user to utilize the bandwidth under
the same conditions as those enclosed in the offer. Once the user
finds an offer (and associated credential) that is acceptable, she
must issue to the ISP a microcheck for this offer.  The microchecks
are encoded as KeyNote credentials that authorize payment for a
specific transaction.  The user creates a KeyNote credential signed
with her public key and sends it, along with her credential from
Provisioning, to the first network element of the ISP. This credential
is effectively a check signed by the user (the Authorizer) and payable
to the ISP (the Licensee).  The conditions under which this check is
valid match the offer sent to the user by the ISP.  Part of the offer
is a nonce, which maps payments to specific transactions, and prevents
double-depositing of microchecks by the ISP.

To determine whether he can expect to be paid (and therefore whether
to accept the payment), the ISP passes the action description (the
attributes and values in the offer) and the user's key along with the
ISP's policy (that identifies the Provisioning key), the user
credential (signed by \bx\ ), the offer credential (signed by the
ISP), and the microchecks credential (signed by the user) to his local
KeyNote compliance checker.  If the compliance checker authorizes the
transaction, the ISP is guaranteed that Provisioning will allow
payment.  The correct linkage among the Merchant's policy, the
Provisioning key, the user key, and the transaction details follow
from KeyNote's semantics \cite{KeyNote}. If the transaction is
approved, the ISP can configure the appropriate routers such that the
user's traffic is treated according to the offer, and store a copy of
the microcheck along with the user credential and associated offer
details for later settlement and payment.

Periodically, the ISP will `deposit' the microchecks (and associated
transaction details) he has collected to the Clearing and Settlement
Center (CSC).  The CSC may or may not be run by the same company as
the Provisioning, but it must have the proper authorization to
transmit billing and payment records to the Provisioning for the
customers.  The CSC receives payment records from the various ISPs;
these records consist of the offer, and the KeyNote microcheck and
credential from the user sent in response to the offer. In order to
verify that a microcheck is good, the CSC goes through a similar
procedure as the ISP did when accepting the microcheck.  If the
KeyNote compliance checker approves, the check is accepted.  Using her
public key as an index, the user's account is debited for the amount
of the transaction.  Similarly, the ISP's account is credited for the
same amount.

\SubSection{\bx\ Operation}
\label{ss:ops}

Having seen the overall system architecture, let us look at a
particular example. {\em Alice} is a user who wants to reserve some
bandwidth for a particular link with {\em Nick's} ISP.  Every evening
Alice contacts her banker and obtains a fresh {\em Check Guarantor}
credential, which allows her to issue KeyNote microchecks. The CG
credential (most of the hex digits from the keys have been removed for
brevity) allows Alice to write checks for up to 5 US Dollars, and she
can do so until March 24th, 2004.

\setlength{\credwidth}{\textwidth}
\addtolength{\credwidth}{-.1\textwidth}

\begin{center}
\renewcommand{\baselinestretch}{1}
\fbox{{\small
\begin{minipage}{\credwidth}
\begin{tabbing}
{\tt \ \ \ \ \ \ }\=\ \kill
{\tt Keynote-Version: 2} \\
{\tt Local-Constants: }\\
\>{\tt ALICE\_KEY = "rsa-base64:MCgCIQ... } \\
\>{\tt CG\_KEY = "rsa-base64:MIGJAo..."} \\
{\tt Authorizer: CG\_KEY} \\
{\tt Licensees: ALICE\_KEY} \\
{\tt Conditions: app\_domain == "Band-X" \&\& }\\
\>{\tt currency == "USD" \&\& \&amount < 5.01} \\
\>{\tt \&\& date < "20040324" -> "true"; } \\
{\tt Signature: "sig-rsa-sha1-base64:QU6SZ..."}
\end{tabbing}
\end{minipage}
}}
\end{center}

Alice now wants to reserve some bandwidth to Dublin. She searches the
\bx\ for a suitable offer, and locates one issued by Nick's ISP that
contains the following Offer credential, indicating that she could
purchase 50Mbps on the specific link (``Dublin-NYC'') for 3 US
dollars:

\begin{center}
\renewcommand{\baselinestretch}{1}
\fbox{{\small
\begin{minipage}{\credwidth}
\begin{tabbing}
{\tt \ \ \ \ \ \ }\=\ \kill
{\tt Keynote-Version: 2} \\
{\tt Local-Constants: }\\
\>{\tt ISP\_KEY = "rsa-base64:7231f..."} \\
{\tt Authorizer: ISP\_KEY } \\
{\tt Licensees: } \\
{\tt Conditions: app\_domain == "Band-X" \&\& } \\
\>{\tt currency == "USD" \&\& }\\
\>{\tt \&bandwidth <= "50Mbps" \&\& } \\
\>{\tt link\_name == "Dublin-NYC" \&\& }\\
\>{\tt \&amount >= 3.00 }\\
\>{\tt \&\& date < "20031120 -> "true"; } \\
{\tt Signature: "sig-rsa-sha1-base64:ab1XXA..."}
\end{tabbing}
\end{minipage}
}}
\end{center}

\noindent Alice then writes a check for the appropriate amount:

\begin{center}
\renewcommand{\baselinestretch}{1}
\fbox{{\small
\begin{minipage}{\credwidth}
\begin{tabbing}
{\tt \ \ \ \ \ \ }\=\ \kill
{\tt Keynote-Version: 2} \\
{\tt Local-Constants: }\\
\>{\tt ALICE\_KEY = "rsa-base64:Mcg..."} \\
\>{\tt ISP\_KEY = "rsa-base64:7231f..."} \\
{\tt Authorizer: ALICE\_KEY} \\
{\tt Licensees: ISP\_KEY} \\
{\tt Conditions: app\_domain == "BAND-X" \&\& }\\
\>{\tt currency == "USD" \&\& amount == "4.25" }\\
\>{\tt \&\& nonce ==  "eb2c3dfc8e9a" \&\& } \\
\>{\tt date == "20041120" -> "true"; } \\
{\tt Signature: "sig-rsa-sha1-base64:Qsd..." }
\end{tabbing}
\end{minipage}
}}
\end{center}

The {\tt \small nonce} is a random number that must be different for
each check, guaranteeing that there will be no double-depositing of
checks.  Alice then sends the Offer credential and the micro-check to
Nick's router using a protocol such as RSVP. Nick receives these
credentials, validates the microcheck to make sure that he will get
paid, and configures the router appropriately.  If the check is not
good, Nick will say so, and refuse to accept the file.  Nick will
verify that he will get paid, and will evaluate the Offer credential
and the microcheck using a simple policy such as:

\setlength{\credwidth}{\textwidth}
\addtolength{\credwidth}{-.1\textwidth}

\begin{center}
\renewcommand{\baselinestretch}{1}
\fbox{{\small
\begin{minipage}{\credwidth}
\begin{tabbing}
{\tt \ \ \ \ \ \ }\=\ \kill
{\tt Keynote-Version: 2} \\
{\tt Local-Constants: }\\
\>{\tt NICK\_KEY = "rsa-base64:7231f..."} \\
\>{\tt CG\_KEY = "rsa-base64:MIGJAo..."} \\
{\tt Authorizer: POLICY} \\
{\tt Licensees: CG\_KEY \&\& NICK\_KEY} \\
{\tt Conditions: }\\
\>{\tt app\_domain == "BAND-X" -> "true";}
\end{tabbing}
\end{minipage}
}}
\end{center}

This policy says that anything that Nick's key {\em and} the Check
Guarantor's key jointly authorize is allowed. Thus, Alice must submit
a valid payment and a valid Offer credential. Since the bandwidth was
paid for, and a path can be found from {\sc POLICY} to a user (Alice)
that has delegated to Nick's key, which in turn has created an
open-access Offer credential, the operation is allowed.  As a matter
of business practice, Nick may require periodic payments from Alice in
order to keep the bandwidth reserved.  Alice must know that and send
microchecks at the appropriate intervals.

If additional routers need to be configured in Nick's ISP, the first
router forwards the necessary information to the next. Note that it is
not necessary for the router itself to perform the signature
verifications and policy validations: it can simply refer these
operations to a Policy Decision Point (PDP), as is envisioned by the
IntServ architecture.

\SubSection{Security Analysis}
\label{sec:secan}

Similar to \cite{sodapaper} and \cite{fileteller}, our system has
three types of communication: provisioning, reconciliation, and
transaction. Although delegation of credentials (and thus access
rights to reserved bandwidth) is possible, we do not consider it in
this paper. We shall not worry about any value transfers to banks, as
there already exist well-established systems for handling those. All
communications between \bx, ISPs, and users can be protected with
existing protocols such as IPsec or TLS. This covers both provisioning
and reconciliation, which occur off-line from the actual bandwidth
reservation and use.  Furthermore, the transactions themselves
(establishing the QoS pipes, or the right to use existing pipes) can
be protected through the same means; the only requirement is that the
user can authenticate with each ISP.

The confidentiality of the transmitted data itself is not within the purview
of our system, nor is it a responsibility of the ISP; if the users do
not trust the network with respect to data confidentiality or
integrity, they should use end-to-end security protocols, {\it e.g.,}
IPsec or TLS. We do not impose any limitations that would preclude the
use of these protocols.

The user needs to ensure that the ISPs provide the promised
service. This can be easily verified by the user using a number of
existing protocols and tools. Protecting against over-charging ISPs is
also straightforward: the details of each transaction can be verified
at any point in time, by verifying the credentials and the
offer. Since only the user can create microchecks, a dispute claim can
be resolved by ``running'' the transaction again. Thus, the user is
safe even from a collusion between any number of ISPs and the \bx\
service. The ISP must ensure that they are paid for the services
offered. Since it has a copy of all transactions (the \bx\ credential,
the microcheck, and the offer), it can prove to the \bx, or any other
party, that a transaction was in fact performed.

The \bx\ also needs to be paid for the services offered. Since the
\bx\ does the clearing of the microchecks, the ISP has to provide the
transaction logs to the \bx. The \bx\ can then verify that a
transaction was done, and at what value. A collusion between the ISP
and a user is somewhat self-contradicting: the user's goal is to
minimize cost, while the ISP's is to maximize revenue, each at the
expense of the other. The function of the \bx\ is to verify each
transaction (perhaps sampling, for very large numbers of
transactions), debit the ISP and credit the user (presumably keeping
some commission or small fee in the process): if the ISP does not give
any credentials to the \bx, then no work was done as far as the \bx\
is concerned (and no payments are made, which benefits the user);
claiming more transactions than really happened is not in the best
interest of the user (so no collaboration could be expected in the
direction), and the ISP cannot ``fabricate'' transactions. Since value
is not stored in either the ISP or the user, only a reliable log of
the transactions is needed at the ISP (and, optionally, at the user).

%% file: 90-related.tex
%
%

\Section{Related Work}
\label{sec:rel}

Despite the ever increasing use of time-sensitive protocols ({\it
e.g.,} VoIP, audio on demand, {\it etc.}) bandwidth reservation has
not been particularly successful. This is caused mainly by the
fear that since these applications have modest bandwidth requirements
the operation of a reservation and payment infrastructure would not
be feasible economically.
Recently, however, newer applications such as video on demand, tele-presence,
and Grid Computing, have bandwidth requirements that may constitute
a significant portion of the available bandwidth. In such cases the
overheads associated with the reservation and billing are smaller
(because we are dealing with fewer more expensive reservations), while
the benefits are greater because of the impact of the data flows on
the infrastructure. 

In Grid Computing in particular, efforts are already underway
\cite{herve-uclp}, to allow end-users to create end-to-end light paths
(optical links that allow unstructured access to the fiber
infrastructure) by combining individual segments very much as we
described in the introduction. The current systems, however, are
targeted towards the academic community and hence assume that
end-users have the required expertise and have non-competitive usage
strategies.  Specifically under the ``User Controlled Light Paths''
framework \cite{herve-uclp}, $(a)$ end-users have to be known by the
system in advance, $(b)$ policy enforcement is not addressed, $(c)$
there is no purchasing of bandwidth, since the network is considered a
common resource. In a commercial environment, a similar system must
deal with billing ({\it i.e.,} how the reserved bandwidth can be paid
by the user) and must support bandwidth reservation in a scalable and
secure manner.

\comment{
	Nowadays, with newer applications such as video on demand,
	tele-presence, and Grid Computing, the unit of allocation is large
	enough to allow a smaller number of higher value transactions that
	place reasonable demands on the reservation and payment components
	of a reservation system. Such a system must deal with billing
	({\it i.e.,} how the cost of the reserved bandwidth can be paid
	by the user) and must support a reservation protocol such as RSVP
	that can perform bandwidth reservation in a scalable and
	secure manner. 
}

\noindent {\bf Billing}  \hskip .1in
Internet telephony (or voice over IP) is widely considered to be the
``killer'' application that will convince users that they need QoS
(and the higher prices this implies). This is underlined by the fact
that the literature concentrates on QoS for VoIP applications. Systems
such as OSP \cite{etsi} provide a way for large organizations to
settle payments related to VoIP call clearing. Although OSP is very
close to \bx, it does not involve the end-user, but instead
concentrates on the ISPs. For example OSP only exchanges Call Detail
Records, the ISPs are responsible for handling customer billing and
payment. In other words the model is that of the traditional TELCO
whereby payment is handled either via prepaid cards, or monthly
telephone bills. \bx\ is not bound to a particular signaling mechanism
(such as H.323) and provides far greater flexibility in that users
that have no prior relationship with an ISP can use the reservation
protocol and pay for their bandwidth.
Although many papers have been written on market-based routing, these
are concerned with the use of market-based techniques in routers,
ignoring the problems of accounting, billing and payment. \bx\ can use
any router that supports a reservation protocol (and the \bx\
extensions).

\noindent{\bf Secure QoS Reservations} \hskip .1in
A secure reservation protocol is required to provide a number of
assurances including $(a)$ that only authorized users can make
reservations, $(b)$ that a reservation made by a user can be traced
back to that user, and $(c)$ that users cannot make reservations over
their allocated quota. These are to protect against starvation or,
perhaps even worse, denial of service that can occur when multiple
unauthorized requests result in the allocation of all available
bandwidth thus preventing legitimate users from reserving bandwidth.
The above considerations imply some authentication mechanism and the
use of integrity checks on the transmitted data. OSP runs over TLS
which encrypts the exchanged data. X.509 certificates are used to
authenticate both ends of a transaction. However, this secure
communication is used only for the data exchange between the ISP nodes
running OSP.  Customer identification is still handled via a separate
system that is operated by the ISPs and usually involves some kind of
PIN or password authentication.
In \cite{billing-pricing} the actual charging is delegated to a
``payment-agent'' that is assumed to run on the same machine as the
user. However, no details are provided on how the ``payment-agent''
effects payment.

All the systems we have looked at assume that the user trusts some
provider who determines the cost of the connection. No system tries to
empower the user by providing choice. \bx\ allows the user to select
the best (as defined by the user) providers to handle the connection
and makes sure that at the end of the day everybody gets paid. This
approach is far superior to the piecemeal approaches found in the
literature.

\noindent{\bf Scalability}\hskip .1in
Each reservation carries with it some overhead. This includes both
protocol overhead, but also state that must be maintained by routers
for each reservation. As the number of reservations increases so does
the overhead. Unless there is some kind of aggregation of requests
this overhead will ultimately define an upper bound on the number of
reservations that can be accommodated by the existing infrastructure.
The complexity of some of the proposed systems ({\it e.g.,}
\cite{jarrett-operational}, and \cite{billing-pricing}) and the small
scale of their test-beds ({\it e.g.,} 200 nodes in
\cite{hao-scalability}) casts grave doubts on their ability to scale
to millions of users and thousands of network elements.
Various techniques that attempt to improve scalability through
aggregation are vulnerable to abuse. For example, in \cite{rsvp} the
authors describe request aggregation whereby multiple requests are
merged into a single larger request for the total bandwidth asked for
by the individual requests. This approach, however, may result in an
upstream node declining the single request thus denying access to all
the requests, even through some of the individual requests could have
gone through \cite{killer-res}.

Since \bx\ covers both reservation and payment, the problem of
scalability has to be addressed in both areas. As far as reservation
is concerned, \bx\ uses the RSVP protocol and so can take advantage of
the optimizations and efficiencies that have either been integrated,
or are being considered for inclusion into the protocol. In the area
of billing, the use of the Keynote-based micro-payment architecture has
been shown to scale well \cite{sodapaper}.

%% file: 95-conclusions.tex
%
%

\Section{Summary and Concluding Remarks}
\label{sec:conc}

To minimize network congestion which can cause complaints and
dissatisfaction among users, ISPs overprovision their networks
\cite{currence-causal}.  Unfortunately, unused bandwidth is wasted
since it cannot be saved for later use. While bandwidth remains cheap,
the ISPs can continue to add capacity ahead of the actual demand, but
this state of affairs will only last as long as users of
time-sensitive services prefer the telephony network. The enormous
cost difference between the telephony network and the Internet
provides an implicit subsidy. However, as users switch to the Internet
for their time-sensitive services, ISPs will no longer be able to
expand their networks. We believe that the framework described in this
paper offers a migration path for both users and ISPs through the
creation of an open market for bandwidth over the Internet.  The
reason is that the \bx\ framework supports a competitive market
offering transparency, and security.  At the same time the low
overheads of the \bx\ framework ensure scalability through the use of
a micro-payment environment.

The benefits offered by \bx\ include:
$(a)$ ``instant'' purchases of bandwidth and advanced purchases
allowing the ISPs to plan ahead their resource allocation strategies,
while being able to auction off unused capacity rather that letting
it go at Best-Effort prices,
$(b)$ efficiency, requiring only a few exchanges
between a buyer and sellers to effect a reservation. Moreover,
the use of the Keynote-based micro-payment framework provides
system-wide efficiency and scalability,
$(c)$ compatibility with existing standards: by utilizing an
existing reservation protocol (RSVP), a \bx\ system may be
be deployed with minimum disruption.
$(d)$ trades between parties that have no established business relationships:
The Credit Institution(s) link buyers and sellers, thus allowing
a transaction to go through without the need for a buyer to be known
to the seller. This is a key requirement for the bandwidth market to
work freely with the buyer being able to select the seller offering
the best value for money.
$(e)$ openness: the \bx\ model allows the presence of multiple
entities for each role ({\it i.e.,} we can have multiple Credit
Institutions, Clearing Houses, buyers and sellers) operating within
a single market. This increases the competition and overall reliability
of the entire system.

%% file: bandex.bbl
\begin{thebibliography}{10}\setlength{\itemsep}{-1ex}\small

\bibitem{diffserv:rfc}
S.~Blake, D.~Black, M.~Carlson, E.~Davies, Z.~Wang, and W.~Weiss.
\newblock {An Architecture for Differentiated Services}.
\newblock Technical report, IETF RFC 2475, December 1998.

\bibitem{KeyNote}
M.~Blaze, J.~Feigenbaum, J.~Ioannidis, and A.~D. Keromytis.
\newblock {The KeyNote Trust Management System Version 2}.
\newblock Internet RFC 2704, September 1999.

\bibitem{sodapaper}
M.~Blaze, J.~Ioannidis, and A.~D. Keromytis.
\newblock {Offline Micropayments without Trusted Hardware}.
\newblock In {\em Proceedings of the Fifth International Conference on
  Financial Cryptography}, 2001.

\bibitem{rfc-rsvp}
R.~Braden, L.~Zhang, S.~Berson, S.~Herzog, and S.~Jamin.
\newblock {Resource {R}e{S}er{V}ation Protocol {(RSVP)} -- Version 1 Functional
  Specification}.
\newblock Internet RFC 2208, 1997.

\bibitem{davie2003}
L.~Burgstahler, K.~Dolzer, C.~Hauser, J.~J\"ahnert, S.~Junghans, C.~Maci\'an,
  and W.~Payer.
\newblock {Beyond Technology: The Missing Pieces for QoS Success}.
\newblock In {\em Proceedings of the ACM SIGCOMM Workshop on Revisiting IP QoS
  (RIPQOS), held in conjunction with the ACM SIGCOMM conference}, August 2003.

\bibitem{currence-causal}
M.~Currence, A.~Kurzon, D.~Smud, and L.~Trias.
\newblock {A Causal Analysis of Usage-Based Billing on IP Networks}, 2000.

\bibitem{payer2003}
B.~Davie.
\newblock {Deployment Experience with Differentiated Services}.
\newblock In {\em Proceedings of the ACM SIGCOMM Workshop on Revisiting IP QoS
  (RIPQOS), held in conjunction with the ACM SIGCOMM conference}, August 2003.

\bibitem{billing-pricing}
R.~J. Edell, N.~McKeown, and P.~Varaiya.
\newblock {Billing Users and Pricing for {TCP}}.
\newblock {\em {IEEE} Journal on Selected Areas in Communications},
  13(7):1162--1175, 1995.

\bibitem{etsi}
ETSI.
\newblock {Telecommunications and Internet Protocol Harmonization Over Networks
  (TIPHON): Inter-domain pricing, authorisation, and usage exchange },
  booktitle = { ETSI DTS/TIPHON-03004 V1.3.0 (1998-09)}, 1998.

\bibitem{herve-uclp}
H.~Guy.
\newblock {Everything you ever wanted to know before you use and/or deploy UCLP
  on your advanced network}.
\newblock In {\em {Proceedings of the CANARIE's Advanced Networks Workshop}},
  November 2004.

\bibitem{hao-scalability}
F.~Hao.
\newblock {Scalability Techniques in QoS Routing}, 2000.

\bibitem{fileteller}
J.~Ioannidis, S.~Ioannidis, A.~Keromytis, and V.~Prevelakis.
\newblock {Fileteller: Paying and Getting Paid for File Storage}.
\newblock In {\em Proceedings of the Sixth International Conference on
  Financial Cryptography}, March 2002.

\bibitem{killer-res}
M.~Talwar.
\newblock {RSVP Killer Reservations, IETF Draft (draft-talwar-rsvp-kr-01.txt)},
  1999.

\bibitem{jarrett-operational}
W.Jarrett, T.Michalareas, and L.Sacks.
\newblock {Operational Support Issues for IP QoS Based Networks}.
\newblock In {\em Proceedings of the IEE Services over the Internet
  Colloquium}, June 1999.

\bibitem{rsvp}
L.~Zhang, S.~Deering, and D.~Estrin.
\newblock {RSVP}: {A} new resource {ReSerVation} protocol.
\newblock {\em {IEEE} network}, 7(5), September 1993.

\end{thebibliography}
